\documentclass[aps,prl,twocolumn,superscriptaddress,showpacs]{revtex4}
\usepackage{placeins}
\usepackage{graphicx}
\usepackage{amsmath}
\usepackage{amssymb}
\usepackage{times}
\usepackage{color}

\begin{document}
\title{Co-evolution of strategy and structure in complex networks \\ 
with dynamical linking}
\author{Jorge M.\ Pacheco}
\affiliation{Program for Evolutionary Dynamics, Harvard University, Cambridge MA 02138, USA}
\affiliation{Centro de F{\'\i}sica Te{\'o}rica e Computacional, 
             Departamento de F{\'\i}sica da Faculdade de Ci{\^e}ncias, 
             P-1649-003 Lisboa Codex, Portugal}
\author{Arne Traulsen}
\affiliation{Program for Evolutionary Dynamics, Harvard University, Cambridge MA 02138, USA}
\author{Martin A.\ Nowak}
\affiliation{Program for Evolutionary Dynamics, Harvard University, Cambridge MA 02138, USA}
\affiliation{Department of Organismic and Evolutionary Biology, Department of Mathematics, Harvard University, Cambridge, MA 02138, USA}

\date{ \today}

\begin{abstract}
Here we introduce a model in which individuals differ in the rate at which they seek new interactions with others, making rational decisions modeled as general symmetric two-player games. Once a link between two individuals has formed, the productivity of this link is evaluated. Links can be broken off at different rates. 
We provide analytic results for the limiting cases where linking dynamics is much faster than evolutionary dynamics and vice-versa, 
and show how the individual capacity of forming new links or severing inconvenient ones maps into the problem 
of strategy evolution in a well-mixed population under a different game. 
For intermediate ranges, we investigate numerically the detailed interplay determined by 
these two time-scales and show that the scope of validity of the analytical 
results extends to a much wider ratio of time scales than expected. 
\end{abstract}

\maketitle

Networks pervade all sciences
\cite{BAreview,ZFMbook,newman2003,AmaralPNAS,May2006}. 
During recent years, researchers have developed methods to characterize such networks,
providing novel insights into the properties accruing to those networked systems
and organizations. 
The classical social network metaphor \cite{SollaPrice} places individuals at the nodes of a network, the network links  representing 
interactions or connections between those individuals. Citation networks, collaboration networks, 
co-authorship and movie co-acting networks, as well as the networks of 
sexual relations, all fall into this metaphoric representation \cite{BAreview,ZFMbook,newman2003,AmaralPNAS,May2006}. 
Most analytical studies on this type of networks carried out to date have aimed to explain the 
emergence of the observed topological properties, as deduced from the empirical data. Networks, however, are dynamical entities, and in this sense the empirical information often only 
provides a fixed-time snapshot of networks which are continuously evolving.
Furthermore, dynamical features of networks have been 
studied in connection with their growth, 
modeled in terms of the preferential attachment 
(or cumulative advantage) mechanism \cite{Simon,SollaPrice,BAreview}, via random addition and removal of nodes \cite{ZFMbook} or by 
imposing different forms of connectivity saturation 
\cite{AmaralPNAS,Mendes2000,BarPRL2000,Holme2006}. 
Moreover, individual decisions to establish or remove/rewire a given link have been studied 
by numerical simulations 
\cite{AmaralPNAS,Mendes2000,BarPRL2000,Holme2006,CoEvol1,CoEvol2,CoEvol3,CoEvol4,Holme2006b}. 
Here we develop a new model which incorporates decisions of individuals when establishing new links or 
giving up existing links. Individuals are capable of making rational choices, modeled in terms of a game, 
associated with well defined strategies. 
We use evolutionary game theory \cite{tj78,ms82,hs1998,hs2003,ns2004} and study the dynamical co-evolution of individual strategies and network structure. We restrict our analysis to symmetric two-player games, 
although the model can be easily extended to games with an arbitrary number of strategies. 
We obtain analytical results which are formally valid in the extreme limits when one of the dynamics (strategy or structure) 
dominates the other, although our numerical simulations show that the range of validity of the analytical results is much wider. 
The present model leads to {\it single-scale} networks as defined in \cite{AmaralPNAS}, with associated cumulative degree distributions 
exhibiting fast decaying tails \cite{AmaralPNAS}, as shown in Fig.~1. 
Such  tails which decay exponentially or faster than exponential, leading to what are known 
as ``broad-scale" and ``single-scale" networks, respectively \cite{AmaralPNAS}, are features which, 
together with a large variability in the average connectivity \cite{BAreview,ZFMbook,May2006}, 
characterize most real-world social networks \cite{newman2003,AmaralPNAS}.
We start by characterizing the networks emerging from our model. 
Subsequently, we introduce individuals who adopt definite strategies and make rational decisions by 
engaging in a game with others, 
studying how strategy and structure co-evolve. Finally, we strenghten the coupling between strategy and structure by 
letting individuals evaluate the productivity of links in which they participate. 

Let us first consider the structural evolution in a population of two types of individuals (players), $A$ and $B$, occupying the nodes of a network. 
The total population size is constant $N=N_A+N_B$. 
Links define interactions between individuals, being formed at certain rates and having specific life-times. 
The maximum possible number of $AA$, $AB$ and $BB$ links is respectively given by  
$N_{ij} = N_i(N_j-\delta_{ij})/(1+\delta_{ij})$ ($i,j= \{A,B \}$).
Suppose $A$ and $B$ players have a propensity to form new links denoted by $\alpha_A$ and $\alpha_B$, such that $ij$ links are 
formed at rates $\alpha_i \alpha_j$. 
The death rates are given by $\gamma_{ij}$ (with associated lifetimes $\tau_{ij}=\gamma_{ij}^{-1}$).
With these definitions the mean field equations governing 
what we call the Active Linking (AL) dynamics of this network are
\begin{equation}
\dot X_{ij} = \alpha_i \alpha_j (N_{ij}-X_{ij}) - \gamma_{ij} X_{ij}. 
\end{equation}
These differential equations lead to an equilibrium distribution of links given by 
$X_{ij}^*=N_{ij}\phi_{ij}$, where $\phi_{ij}=\alpha_i \alpha_j (\alpha_i \alpha_j + \gamma_{ij})^{-1}$
denotes the fraction of active links.
In Fig.~1 we show how this model leads to stationary regimes of complex networks which can exhibit different degrees of heterogeneity.
In particular, the the dependence of the stationary networks on the frequency of individuals of a given type
will automatically couple network dynamics with the frequency-dependent evolutionary dynamics we introduce in the following. 

Let us now introduce a game between $A$ and $B$ 
leading to frequency-dependent evolution of strategies.
The game is given by the payoff-matrix $M_{ij}$
\[
\bordermatrix{
  & A & B \cr
A & a & b \cr
B & c & d \cr}. 
\]
In the stationary regime of Active Linking (AL) dynamics, the average fitness of $A$ and $B$ individuals is given by 
$f_i = \sum_j M_{ij} \phi_{ij} (N_j- \delta_{ij})$.
It is noteworthy that these expressions are equivalent to the average payoffs of $N_A$ and $N_B$ players who play a game specified by the (rescaled) payoff-matrix $M'_{ij}=M_{ij}\phi_{ij}$ on a complete graph. 

\begin{figure}[hpbt]
\begin{center}
\vspace{1cm}
{\includegraphics[width=8.5cm]{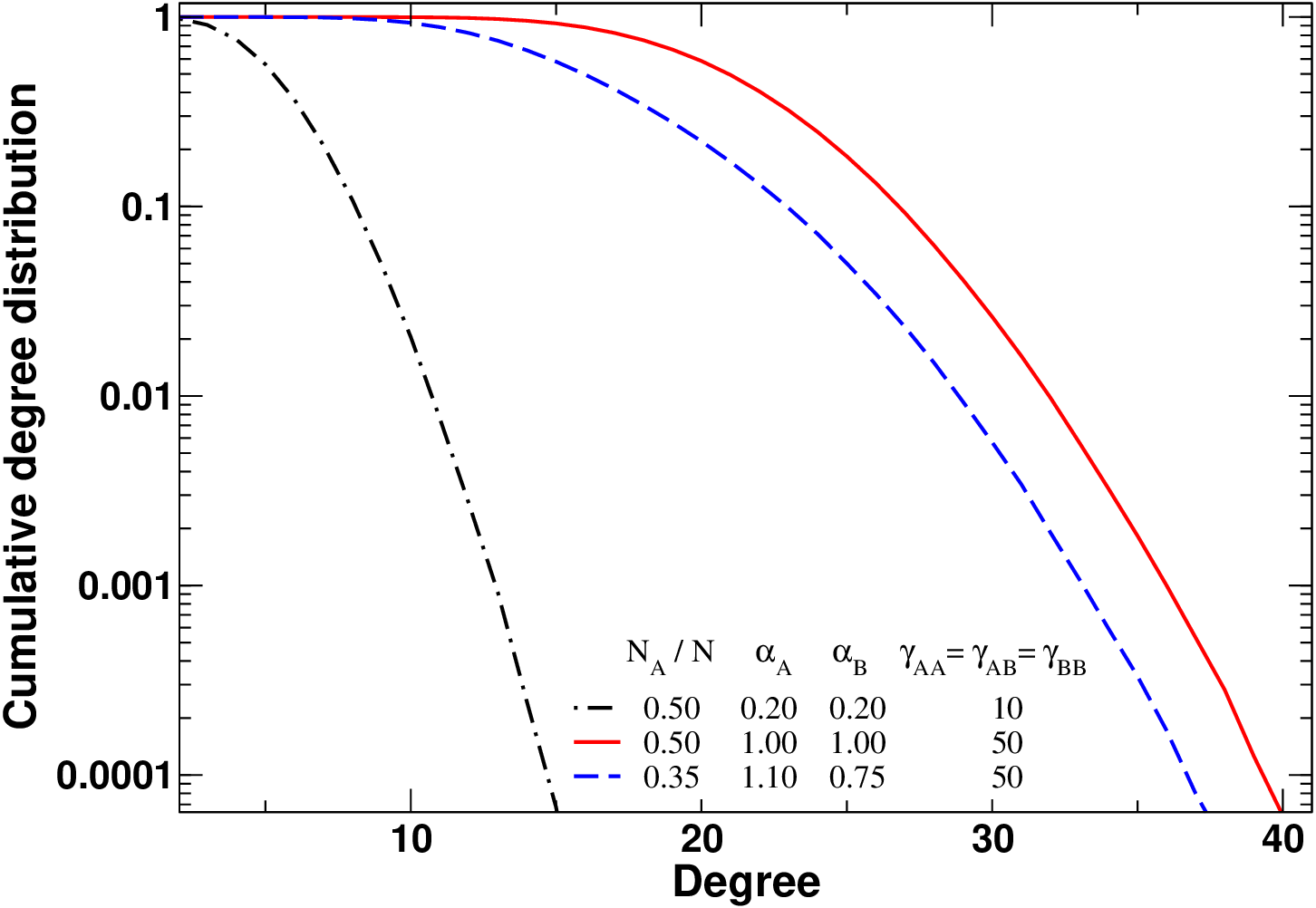}}
\caption{(color online)
Cumulative degree distributions for networks generated with the present model, for populations of size $N=10^3$ 
and two different types of individuals. 
The fast decaying tails correlate well with the observed tails 
of real social networks \cite{BAreview,ZFMbook,newman2003,AmaralPNAS,May2006}. On the other hand, 
the dependence of the final network on the frequency of each type of individuals leads to a 
natural coupling between network dynamics and frequency-dependent strategy evolution. 
}
\end{center}
\end{figure}

So far we have dealt with AL-dynamics. Let us now 
study how the frequencies of strategies $A$ and $B$ change under evolutionary game dynamics. We assume that the characteristic time-scale associated with AL-{\it dynamics} is $T_a$, whereas that associated with 
{\it strategy updating} is $T_s$. 

Reproduction can be genetic or cultural. We adopt the pairwise comparison rule, which provides a convenient framework 
of game dynamics at all intensities of selection \cite{tpn06a}. 
Two individuals from the 
population, $A$ and $B$ are randomly chosen for update. The strategy of $A$ will replace that of $B$ 
with a probability given by the Fermi function $p= [{1+e^{- \beta (f_A-f_B)}}]^{-1}$. The reverse will happen with probability $1-p$.
The quantity $\beta$ 
controls the intensity of selection. 
For $\beta \to \infty$ 
the individual with the lower payoff 
deterministically adopts the strategy of the other individual. 
For 
$\beta \ll 1$, we recover 
the weak selection limit of the frequency dependent Moran process \citep{nowak04}.

When $T_a \ll T_s$, AL proceeds much faster than strategy update on each node. 
Hence, the stationary regime of AL-dynamics determines the average payoff and fitness of individuals. 
This means that strategy evolution proceeds as in a well-mixed population of $A$ and $B$ players 
(complete graph) engaged in a game specified by the payoff-matrix $M'_{ij}=M_{ij}\phi_{ij}$. 
Since AL-dynamics is fast, the dynamics of the system does not depend on the starting condition, and 
we can compute analytically the fixation probabilities of strategies $A$ and $B$.
The probability $\rho_A(k)$ that $k$ $A$-players introduced into a population of 
$B$-players will take over the entire population is given by \citep{tpn06a} 
\begin{equation}
\rho_A(k)= 
{
\frac{{\rm erf}\left[ \xi_k\right]-{\rm erf}\left[ \xi_0\right]}
{{\rm erf}\left[\xi_N \right]-{\rm erf}\left[ \xi_0\right]}
},
\label{rhoa}
\end{equation}
where ${\rm erf}(x)$ is the error function and
$\xi_k=\sqrt{ {\frac{\beta}{u}}} \left(k  u + v \right) $.  
We have $2u = a'-b'-c'+d'$ and $2 v = -a'+b' N -c' N+c'$. 
For $u=0$, Eq.~2 simplifies to 
$\rho_A(k)= {{(e^{-2 \beta v k}  -1)}/{(e^{-2 \beta v N}  -1})}$. 

On the other hand, in the opposite limit where $T_a \gg T_s$, evolution will proceed according to the usual game dynamics 
\cite{NS01,NS02,NS04,NS07,NS08,NS10,NS11,NS12,NS13,NS14,NS15,NS16,NS17,NS18} 
on a {static} network reflecting the initial configuration. 
If we start from a {complete graph} then Eq.~(2) remains valid, except that $u$ and $v$ must be calculated employing the original 
payoff-matrix $M_{ij}$. If we start from another graph topology, 
analytical and numerical results for static networks apply instead \cite{lieberman,ohtsukinat,ohtsukijtb}. 
Whenever $T_a \sim T_s$, one expects a detailed interplay between these two processes to drive 
co-evolution. This regime can be explored by computer simulations of AL-dynamics.
As an example, we investigate the interaction between cooperators and defectors in the 
Prisoner's Dilemma (PD). 
A cooperator, $C$, pays a cost $c$ for every link, and the partner of this link receives a benefit $b>c$. 
Defectors, $D$, pay no cost and distribute no benefits. 
The payoff-matrix becomes
\[
\bordermatrix{
  & C & D \cr
C & b-c & -c \cr
D & b & 0 \cr}.
\label{bcmatrix}
\]
\par\noindent

\begin{figure}[hpbt]
\begin{center}
\vspace{1cm}
{\includegraphics[width=8.5cm]{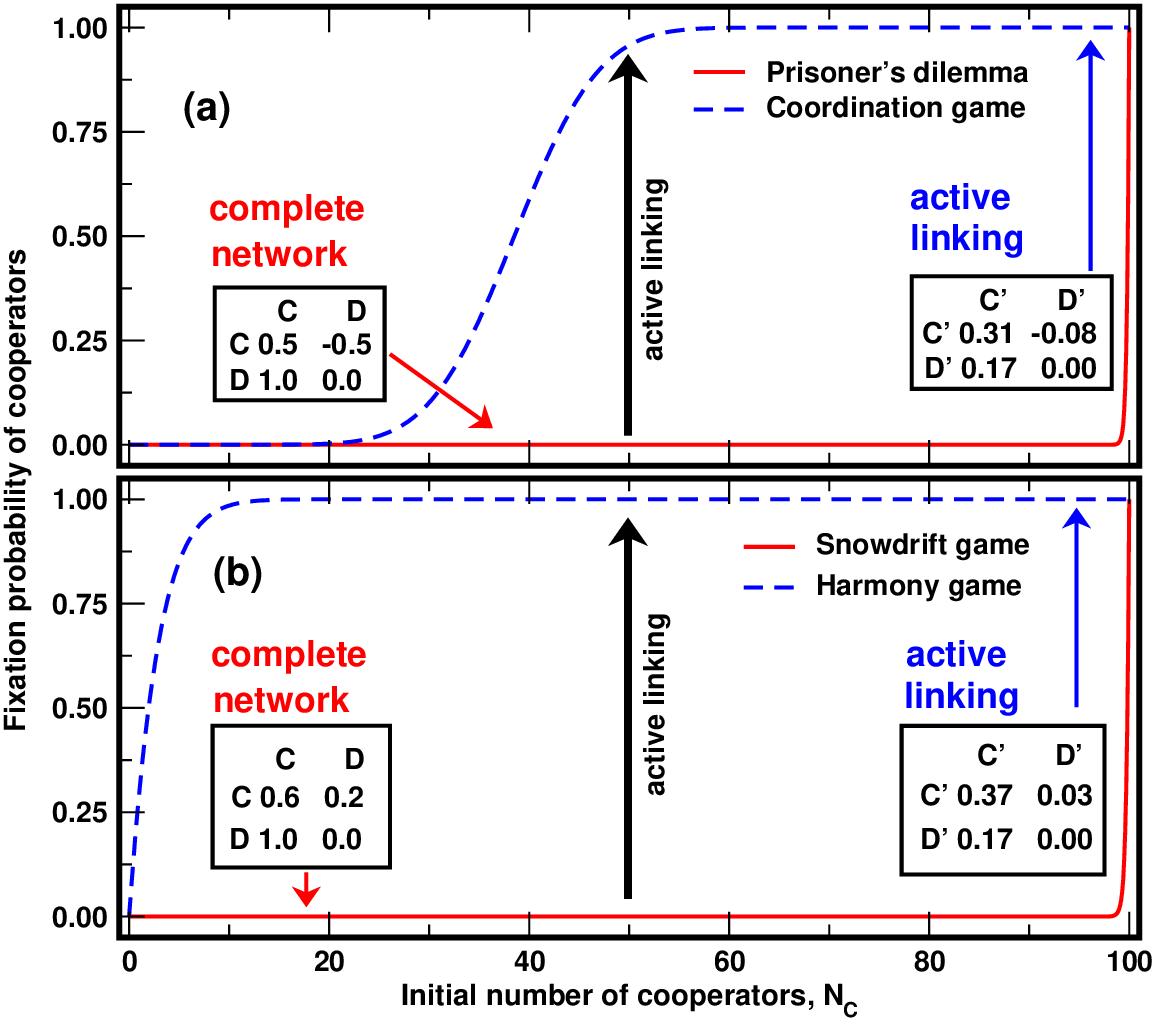}}
\caption{(color online)
Fast active linking (AL) changes the dynamics of the social network and the payoff matrix.
a) AL transforms a Prisoner's Dilemma with $c=0.5$ and $b=1.0$ into a coordination game. While fixation of cooperators is negligible in the Prisoner's Dilemma (solid line), cooperators can take over 
with AL (dashed line) 
b) A snowdrift game with $c=0.8$ and $b=1.0$ in which fixation of defectors is certain is transformed into a harmony game in which cooperators prevail. 
The vertical arrows show how AL affects the fixation probability when initially $50 \%$ 
cooperators are present: while they have no chances on a complete graph despite their high abundance, AL makes fixation of cooperation almost certain in both systems
($\beta=0.1$, $N=100$, $\alpha_C=\alpha_D=0.4$, 
$\gamma_{CC} = 0.1$, 
$\gamma_{CD} = 0.8$, 
$\gamma_{DD} = 0.32$).
}
\end{center}
\end{figure}

On complete graphs, 
cooperators are never advantageous compared to defectors. This means if $T_a \gg T_s$ cooperators are never favored by selection. On the other hand, 
if $T_a \ll T_s$, the effective payoff-matrix is different, and may not correspond anymore to a PD, that is, 
when AL dominates, the problem becomes equivalent to 
the evolutionary dynamics of a different game in a 
complete graph. 
The advantage of cooperators from AL can be captured by the parameter $r=(\phi_{CC}-\phi_{CD})/\phi_{CC}$ 
which provides a measure of the advantage of assortative interactions ($CC$-links) with respect to disassortative ones ($CD$-links). 
In terms of $r$, the PD is transformed into a coordination game
whenever $r>c/b$, which is formally equivalent to Hamilton's rule of kin selection \cite{hamilton}. 
In strategy phase space, an unstable interior fixed point 
develops at a frequency of cooperators given by 
$N_C/N \approx (1-r)c/[r(b-c)]$. 
In other words, for $T_a \ll T_s$ fixation of cooperators is almost certain if the initial fraction of cooperators in the population exceeds this ratio.
In Fig.~2a we provide numerical examples of this scenario, 
whereas in Fig.~3 we investigate the behavior of our co-evolutionary model as $T_s / T_a$ increases. 
The results of Fig.~3 show how the ratio of time scales affects co-evolution of strategy and structure. 
In all cases we start from well-mixed populations (complete graphs). 
Clearly, not only the asymptotic behavior coincides with the analytic prediction but, perhaps more importantly, 
Fig.~3 shows that only for $0.01 \leq T_s / T_a \leq 0.1$ does the interplay between the two time scales deviate significantly 
from the analytic predictions. 

\begin{figure}[hpbt]
\begin{center}
\vspace{1cm}
{\includegraphics[width=8.5cm]{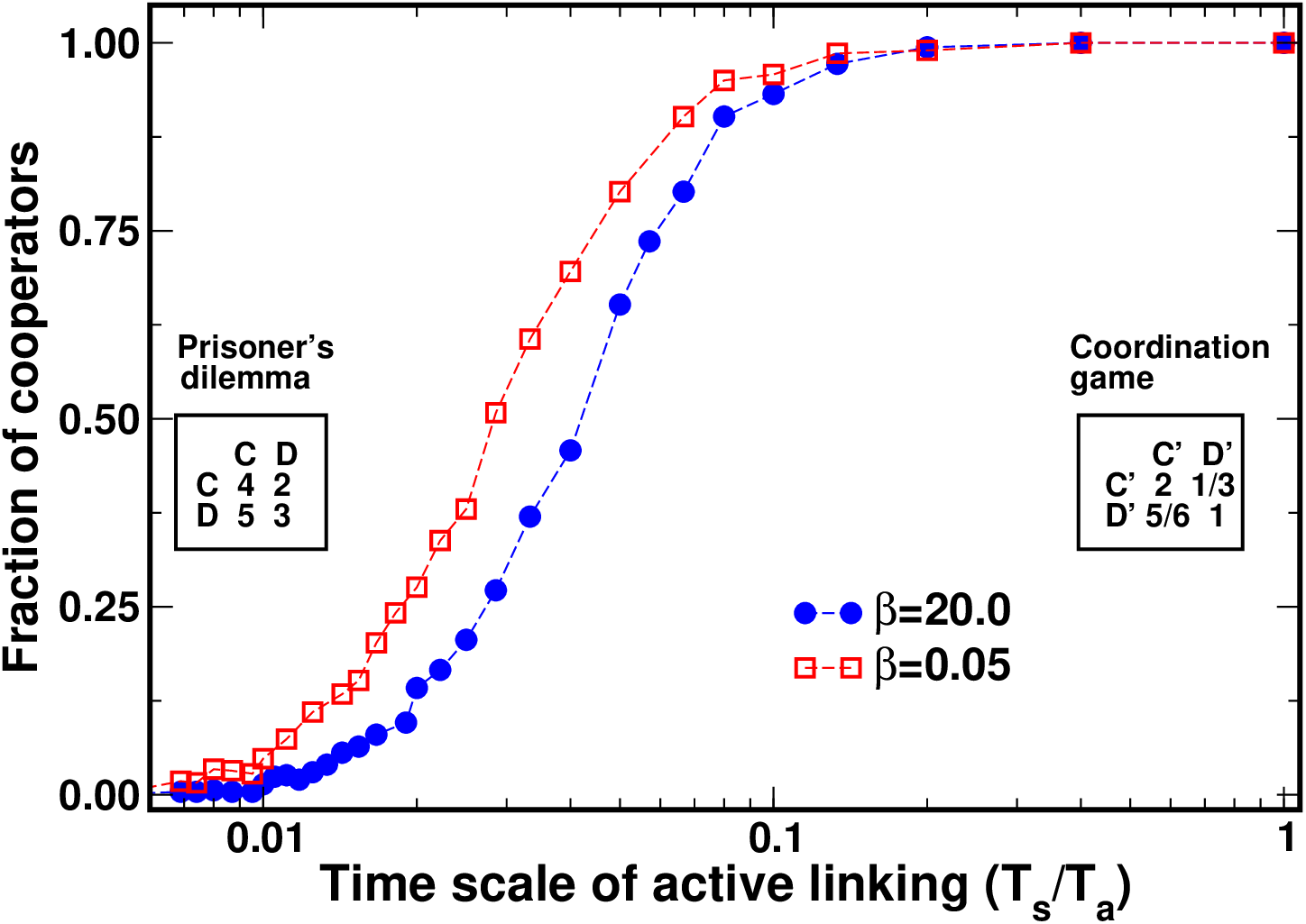}}
\caption{(color online). Co-evolutionary dynamics of strategy and structure. 
The curves drawn correspond to the results of computer simulations carried out for 
networks of size $N=100$. 
Parameters and payoff matrices are the same as in Fig.~2. 
The rescaled payoff-matrix leads to a fixed point at a fraction 
of cooperators $N_C/N \approx 35\%$. 
For each value of the ratio $T_{s}/T_{a}$, we ran 100 simulations, starting from 
$50\%$ cooperators and a complete graph. 
The values plotted correspond to the fraction of runs which ended with $100 \%$ cooperators.  
We fix $T_{s}=1$ and vary $T_{a}$. 
In each time step, synchronous updating of strategies is carried out with probability 
${T_a}^{}/({T_s}^{}+{T_{a}}^{})$
using Fermi-update, AL beeing carried out otherwise.
For the extreme limits we obtain perfect agreement with the analytic predictions. 
However, the analytic results remain valid for a much larger range of values $0.01 \leq T_{a}/T_{s} \leq 0.1$ 
than one would expect from pure theoretical considerations. Indeed, only between these two limits a crossover takes place, which depends on the intensity of selection 
$\beta$ as illustrated. 
}
\end{center}
\end{figure}

Finally, let us further couple the dynamics of links and the dynamics of strategies by introducing payoff dependent AL-dynamics.
An interesting coupling arises when we associate the propensity to form new links
and the lifetime of different types of links with the productivity of those links assessed in terms of payoffs. 
Many possibilities can be readily envisaged, which will lead to different context-based justifications for the choices 
of parameters $\alpha_i$ and $\gamma_{ij}$. Here we explore the case in which 
cooperators and defectors share the same propensity to form new links $\alpha_C = \alpha_D$, whereas 
the lifetimes of $ij$-links are proportional
to the average profit expected from that link. Other linear as well as non-linear alternatives are possible. 
A simple average relation, based on the expected outcome from different types of interactions leads to $\tau_{ij}= (M_{ij}+M_{ji})/2$ which yields $\tau_{CD}=\tau_{CC}/2$, whereas $\tau_{DD}=0$. 
More generally, we may assume that $\tau_{CD}=\tau_{CC}/p$ with $p>1$, maintaining 
$\tau_{DD}=0$ (this results from the zero-entry in the payoff-matrix). 
We may now express $r$ in terms of the constant $\theta=\tau_{CC} \alpha_C^2$, obtaining 
$r=\frac{p-1}{p+\theta}$, an increasing function of $p$. 
The intuition behind this result is clear: 
The larger the value of $p$, that is, the smaller the lifetime of $CD$-links compared to $CC$-links, the smaller the value of $b/c$ above which cooperation will thrive. Moreover, the larger the value of $p$ the smaller the fraction of cooperators that is necessary to be initially present in the population for cooperation to dominate over defection in the resulting coordination game. 

The transformation between a PD and a coordination game is not the only possible one: Inspection of $\phi_{ij}$ shows that other transformations are feasible. 
The Snowdrift Game (SG) has recently attracted a lot of attention, due to its potential biological relevance \cite{NS12}. 
In the SG, a cooperator pays a cost $c$, but two cooperators share this cost. Whenever one player cooperates, both receive a benefit $b>c$, leading to the payoff matrix 
\[
\bordermatrix{
  & C & D \cr
C & b-\frac{c}{2} & b-c \cr
D & b & 0 \cr}.
\]
For strong selection on complete graphs, the SG leads to a stable coexistence between cooperators and defectors, corresponding to a stable interior fixed point 
in strategy phase space at $N_C/N \approx (2b-2c)/(2b-c)$, which becomes small whenever $c \approx b$. Nonetheless, for large $T_{s}/T_{a}$ the SG is effectively transformed 
into the Harmony game, for which 1 cooperator is enough to invade the entire population, see Fig.~2b. 
For the payoff matrix above, the SG is effectively converted into a Harmony game whenever $r>c/(2b)$, where 
$r=(\phi_{CC}-\phi_{CD})/\phi_{CC}$ as above. 
If costs and benefits are the same, the assortment of interactions $r$ has to be only half as high as for the transformation of the PD into a coordination game in order to transform the SG into a Harmony game. 

To sum up, by equipping individuals with the capacity to control the number, nature and duration of their interactions with others, 
we introduce an active linking dynamics which leads to networks exhibiting different degrees of heterogeneity. 
In the limit when active linking dynamics is much faster than strategy dynamics, we obtain a 
simple rescaling of the payoff-matrix. Such rescaling can lead to a transformation of the type of game, effectively taking place in a finite, well-mixed population. 
As the ratio between the time scales associated with linking dynamics and strategy dynamics increases, the interplay between 
these two dynamical processes leads to a progressive crossover between the analytic results discussed here and the evolutionary dynamics of strategies taking place on static graphs. 
Complementing previous numerical explorations \cite{CoEvol1,CoEvol2,CoEvol3,CoEvol4}, the present model provides a simple analytical pathway towards 
understanding of 
how self-interested individuals may actually end up cooperating, 
showing how selective choice of new links (favouring assortative mixing between cooperators) associated with fast rewiring dynamics may 
provide the means to achieve long term cooperation. 

\acknowledgments{
Discussions with K. Sigmund, C. Hauert, H. Ohtsuki and C. Taylor are gratefully acknowledged. 
We acknowledge financial support from FCT, Portugal (J.M.P.), the ``Deutsche Akademie der Naturforscher Leopoldina'' 
(A.T.), the John Templeton Foundation and 
the NSF/NIH joint program in mathematical biology (M.A.N.). 
}

\bibliographystyle{plain}

\end{document}